\documentclass[superscriptaddress,twocolumn,10pt]{revtex4-1}

\usepackage{amsmath}    
\usepackage{graphicx}   
\usepackage{verbatim}   
\usepackage{color}      
\usepackage{subfigure}  
\usepackage{hyperref}   
\usepackage{amssymb}
\usepackage{amsthm}
\usepackage{amsfonts}
\usepackage{lipsum}
\newcommand{\dC}{$^{\circ}$C}

\begin{document}

\title{Elastic Response of Mesoporous Silicon to Capillary Pressures in the Pores}

\author{Gennady Y. Gor} 
\affiliation{NRC Research Associate, resident at the Center for Computational Materials Science, Naval Research Laboratory, Washington, DC 20375, USA}
\email[]{ggor@princeton.edu}
\author{Luca Bertinetti} 
\affiliation{Department of Biomaterials, Max-Planck Institute of Colloids and Interfaces, Research Campus Golm, 14424, Potsdam, Germany}
\author{Noam Bernstein}
\affiliation{Center for Computational Materials Science, Naval Research Laboratory, Washington, DC 20375, USA}
\author{Tommy Hofmann}
\affiliation{Helmholtz-Centre Berlin for Materials and Energy, D-14109 Berlin, Germany}
\author{Peter Fratzl} 
\affiliation{Department of Biomaterials, Max-Planck Institute of Colloids and Interfaces, Research Campus Golm, 14424, Potsdam, Germany}
\author{Patrick Huber} 
\affiliation{Department of Biomaterials, Max-Planck Institute of Colloids and Interfaces, Research Campus Golm, 14424, Potsdam, Germany}
\affiliation{Institute of Materials Physics and Technology, Hamburg University of Technology (TUHH), Ei\ss endorfer Str. 42, D-21073 Hamburg-Harburg, Germany}

\date{\today}

\begin{abstract}
\noindent We study water adsorption-induced deformation of a monolithic, mesoporous silicon membrane traversed by independent channels of $\sim$8 nm diameter. We focus on the elastic constant associated with the Laplace pressure-induced deformation of the membrane upon capillary condensation, i.e. the pore-load modulus.  We perform finite-element method (FEM) simulations of the adsorption-induced deformation of hexagonal and square lattices of cylindrical pores representing the membrane. We find that the pore-load modulus weakly depends on the geometrical arrangement of pores, and can be expressed as a function of porosity. We propose an analytical model which relates the pore-load modulus to the porosity and to the elastic properties of bulk silicon (Young's modulus and Poisson's ratio), and provides an excellent agreement with FEM results. We find good agreement between our experimental data and the predictions of the analytical model, with the Young's modulus of the pore walls slightly lower than the bulk value. This model is applicable to a large class of materials with morphologies similar to mesoporous silicon. Moreover, our findings suggest that liquid condensation experiments allow one to elegantly access the elastic constants of a mesoporous medium.  
 \end{abstract}

\maketitle

Mesoporous silicon (pSi) prepared by electrochemical etching of bulk silicon has been attracting much attention from both fundamental and applied sciences owing to its unique optical, electrical and thermal properties \cite{Canham1990, Goesele1991, Sailor2011, Canham2015, Henstock2015, Huber2015}. While control over mechanical properties of pSi is necessary for its applications, the understanding of its response to mechanical loads is mainly limited to measuring Young's modulus of porous samples \cite{Bellet1996, Populaire2003}.  

Since pSi can be prepared as a bulk (monolithic) mesoporous system with parallel, channel-like independent pores, it has been of a particular interest for studies of fluid adsorption  \cite{Canham2015}. On the one hand, this matrix allows one to study the influence of spatial confinement on the physics and chemistry of liquids \cite{Coasne2002, Wallacher2004, Hofmann2005b, Naumov2008, Kumar2008b, Henschel2009, Karger2014}. On the other hand, it is possible to synthesize composite materials with finely tuned optical and electrical properties by liquid adsorption or infiltration \cite{Westover2014, Canham2015}. Therefore, exploration of the mechanical properties of silicon relevant to interactions with liquids, the topic of this Letter, is also of high importance.

When a fluid is adsorbed in a mesopore, it exerts a pressure on the pore walls, which is typically of the order of $10^7$~Pa \cite{Gor2010}. This pressure causes deformation of the pore, and as a result deformation of the porous material as a whole. This effect, known as adsorption-induced deformation has been experimentally observed for various mesoporous materials: Vycor glass \cite{Amberg1952, Schappert2014Langmuir}, templated silica \cite{Gunther2008, Prass2009, Balzer2015}, low-k films \cite{Mogilnikov2002, Dourdain2008}, aerogels \cite{Reichenauer2001, Herman2006}, porous gold \cite{Shao2010} and pSi \cite{Dolino1996}. There are two ways to measure the adsorption strains experimentally: for materials which can be prepared as macroscopic samples, the dilatometric technique can be used, and the reported strain is the relative change of the length of the sample \cite{Amberg1952, Reichenauer2001, Shao2010, Schappert2014Langmuir}. For crystalline materials or materials which have periodic pore structure, X-ray diffraction can be used, and the strain is calculated as a change of the crystal \cite{Dolino1996, Shao2010} or pore-spacing lattice constant \cite{Gunther2008, Prass2009}. Irrespective of the technique, the measured strains for all these materials (except for aerogels) are of the order of $10^{-4}-10^{-3}$.  Such small strains are in the linear elastic regime, and it is reasonable to assume a linear relation between the pressure in the pore and the experimentally observed strain with a proportionality constant $M$, called the pore-load modulus \cite{Prass2009}. If the pore-load modulus for a certain porous material is known, it is possible to calculate the fluid pressure in the pores based on the experimental data on adsorption-induced strain, providing information on the thermodynamics of the confined fluid. Also, understanding of elastic response of a porous material to adsorption is useful for its application to sensing and actuation \cite{Bertinetti2013, Zhao2014}. 

The pore-load modulus relates the pressure inside the pore to the overall deformation of the sample, rather than relating an external load to a deformation, like Young's or bulk moduli. Therefore an important question is how to relate the pore-load modulus to the material properties of the matrix and the pore geometry. In general, for a material with a wide pore size distribution (PSD) and arbitrary pore morphology and orientation, this question may be complicated. However, for systems such as pSi considered in this work, where the geometry is regular, it can be resolved.  

Here we present a dilatometric study of the deformation of a macroscopic pSi membrane induced by adsorption of water vapor. From the experimental strain isotherm we calculate the pore-load modulus. We perform FEM simulations of adsorption-induced deformation of samples with hexagonal and square lattices of cylindrical pores. We find that the pore-load modulus weakly depends on the geometrical arrangement of pores, and can be expressed as a function of porosity. Therefore we relate the deformation of a porous sample to the deformation of a single cylindrical tube from the pressure applied to its inner surface. Based on this model we derive an analytical expression for the pore-load modulus as a function of the porosity and the elastic properties of the non-porous material (Young's modulus and Poisson's ratio). The predictions of our analytical model are in excellent agreement with FEM results. We also achieve good agreement between our experiment and our analytical model for adsorption-induced deformation of pSi, suggesting that the Young's modulus of silicon pore walls could be only slightly lower than the Young's modulus of bulk silicon.

{Adsorption-induced deformation of pSi has previously been studied by Dolino et al. \cite{Dolino1996}. They obtained the strain normal to the plane of the sample from the shifts of X-ray diffraction peaks of the crystalline lattice of the silicon matrix, and calculated the normal-to-the-plane pore-load modulus from a comparison of the experimental strain to the fluid pressure in the gas pressure region, where the pores are filled with capillary condensate and the fluid pressure in the pore has a simple form. They also proposed an analytical expression for the pore-load modulus as a function of material parameters based on Scherer's model for porous glasses \cite{Scherer1986} and the theory of elasticity for cellular materials \cite{Gibson1999}. However, their model did not agree with the experimental data. Recently Grosman et al. reported a dilatometric study of both the in-plane and normal-to-the-plane deformation of pSi induced by n-heptane adsorption \cite{Grosman2015}. In this work the authors calculated the pore-load moduli for two samples, and proposed their own mechanical model to relate these moduli to the Young's modulus of silicon. By applying their model to their experimental data the authors came to the conclusion that the Young's modulus of silicon walls is lower than the Young's modulus of bulk silicon by a factor of five.}

Another attempt to relate the pore-load modulus to material parameters was made by Prass et al. \cite{Prass2009}. They studied adsorption strains in MCM-41 and SBA-15 silica, which have morphologies similar to pSi, and proposed that the strain, derived from the change of the pore lattice parameter, behaves similarly to the hoop stress in thin-walled cylinders. Although that model provided good agreement with their experimental data, it was not consistent with their finite element method (FEM) calculations. Our model does not employ the thin-wall assumption and is fully supported with FEM calculations.

\begin{figure}[ht]
\centering
\includegraphics[width=0.95\linewidth]{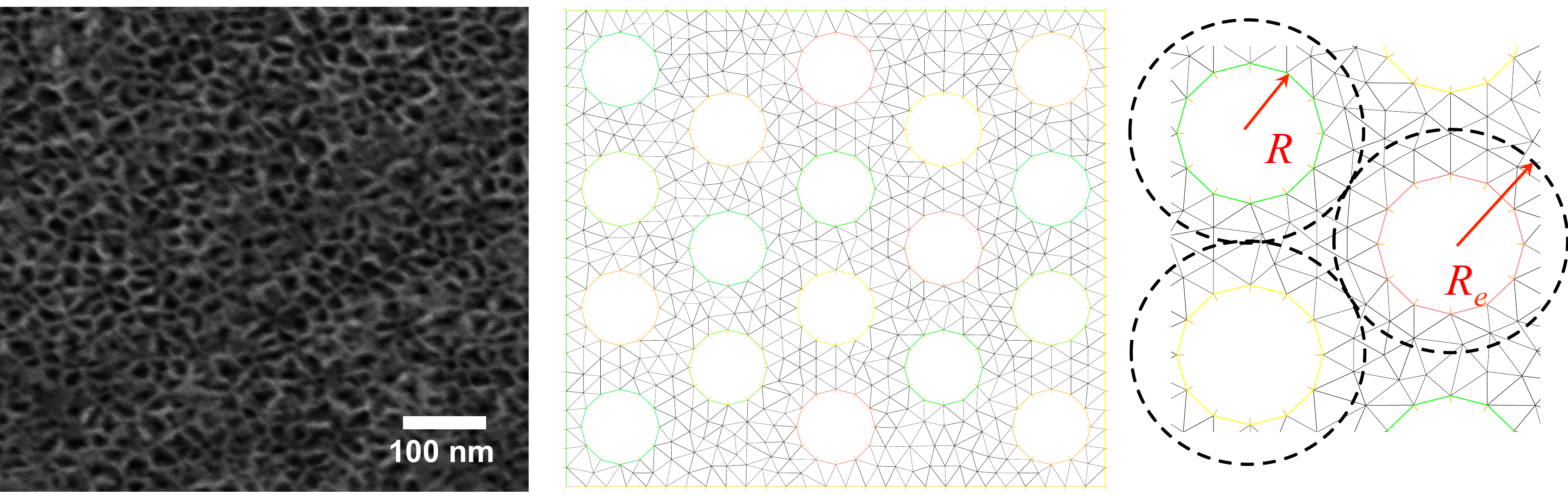}
\caption{SEM image of the pSi structure. Two-dimensonal finite element representation of a porous material consisting of parallel cylindrical pores distributed on a hexagonal lattice. The mesh shown here is much coarser than the mesh used for the FEM calculations. Our FEM analysis shows that the deformation of the porous sample  can be reduced to the deformation of the cylindrical domains in the $R_e$ vicinity of each pore; the latter problem can be solved analytically.}
\label{fig:porous}
\end{figure}

Mesoporous silicon was prepared by electrochemical etching of highly boron-doped (100) silicon wafers (producer: SiMat, Landsberg, Germany; specific conductivity: $\rho$=0.01--0.025 $\Omega$ cm). After an etching depth of 30$\mu$m a high voltage was applied in order to remove the mesoporous part from the bulk silicon underneath. The resulting parallel, non-interconnected, tubular pores have a short-range hexagonal order.  A scanning electron microscope (SEM) image of the sample is shown in Fig.~\ref{fig:porous}.

The sample was cut in a rectangular shape with 4 mm $\times$ 7 mm and a thickness of 30 $\mu$m. The sample's long axis was parallel to the crystallographic $\langle$010$\rangle$ direction. As prepared the inner pore walls are Si-H terminated. A 30 minutes infiltration of a peroxide/water solution and subsequent rinsing with Millipore water rendered the inner pore space Si-OH terminated, and thus hydrophilic.  A volumetric nitrogen adsorption isotherm measurement performed at 77~K and analysed within the NLDFT model \cite{Landers2013} indicated a mean pore diameter of 8~nm (width of the PSD 15\%) and a volume porosity of 60\%.
 
\begin{figure}[ht]
\centering
\includegraphics[width=0.95\linewidth]{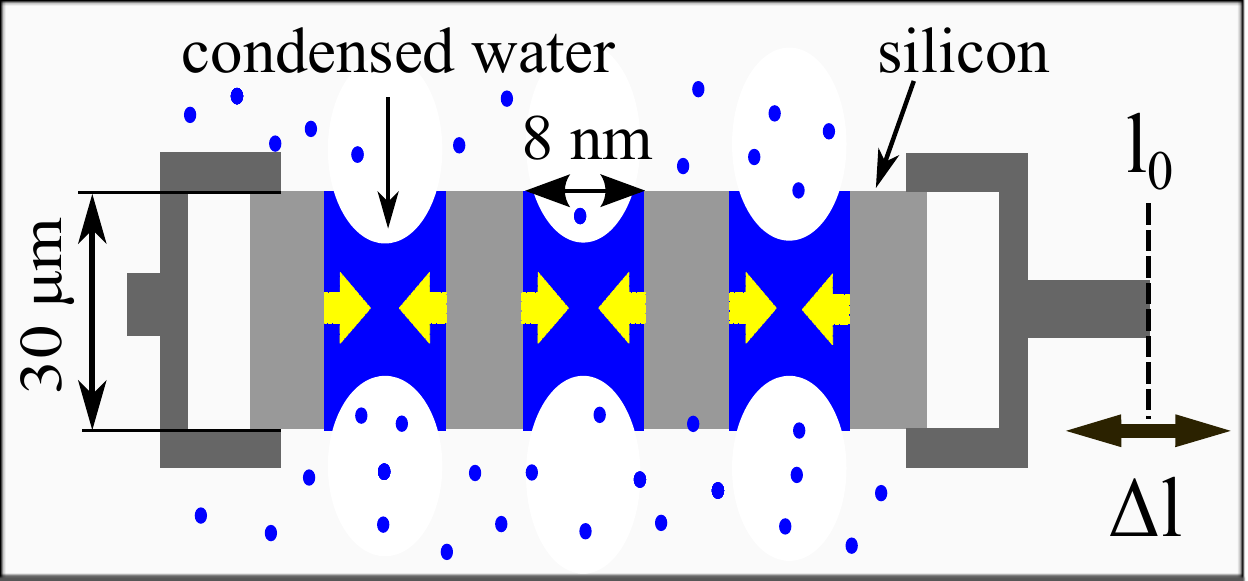} 
\caption{(color online) Schematic of the experimental setup used for measuring the adsorption-induced deformation of porous silicon. The sample holder (dark grey, on the left) is connected to a load cell, and the sample holder on the right is connected to a linear motor stage to measure the length change $\Delta l$. Thick arrows (yellow online) indicate the force on the pore walls due to Laplace pressure upon water condensation.}
\label{fig:setup}
\end{figure} 
 
The measurements of the adsorption-induced strains were performed using a custom-made dilatometric setup, Fig.~\ref{fig:setup}. The samples were tested in a sealed chamber, kept at a constant temperature of 24\dC~by means of a water circulation thermostat (Huber). The humidity inside the chamber was controlled by means of a Wetsys (Setaram) humidity generator, which was working with a flow of 200 ml/min. The samples were clamped to two aluminium holders. The macroscopic deformation was measured by one of the holders that connects to a Physics Instruments M-404 linear motor stage (resolution of 2~$\mu$m), while the axial tensile force was measured using a Honeywell R-30 load cell (50 N max. load), attached to the other holder. The standard deviation of the measured force background noise over more than 100,000 points was 10 mN. The strain of the sample was measured by driving the motors to keep a constant (small) force of 100 mN, while the humidity was continuously changed between 5\% and 95\% RH at a rate of 10\% RH/h. The stress on the sample under these conditions is estimated to be lower than 0.5~MPa. 

\begin{figure}[ht]
\centering
\includegraphics[width=0.95\linewidth]{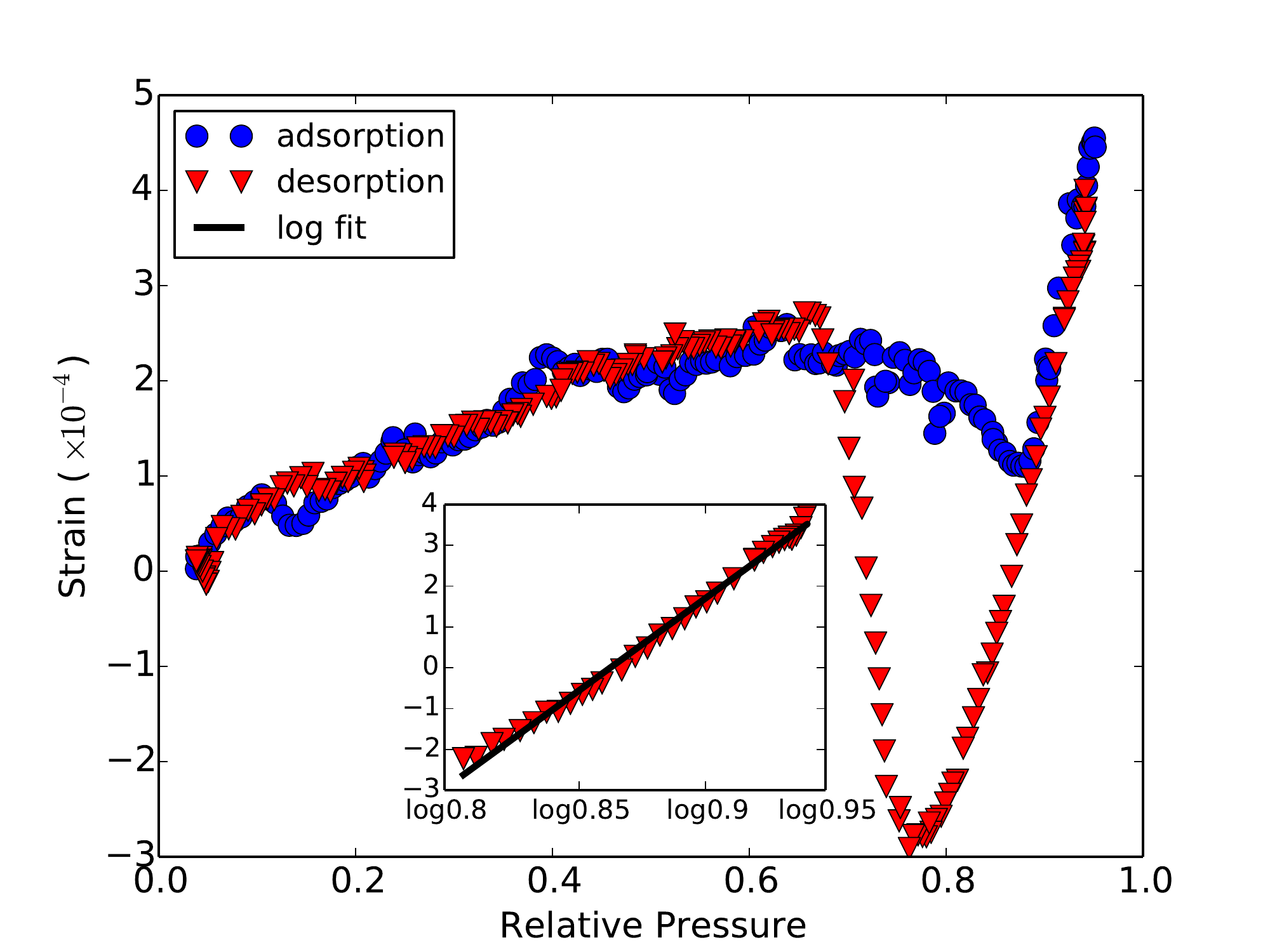} 
\caption{Measured strain of a monolithic mesoporous silicon membrane as a function of relative water vapor pressure $p/p_0$. Inset: Measured strain in the capillary condensation regime, where the strain depends logarithmically on $p/p_0$ and thus linearly on the Laplace pressures in the pores. The solid line represents a linear fit of the strain as a function of log relative pressure, and its slope corresponds to the pore-load modulus $M$.}
\label{fig:expt}
\end{figure}

In Fig.~\ref{fig:expt} we show the experimentally measured, macroscopic strain of the sample $\epsilon_m=\Delta l/l_0$ as a function of relative vapor pressure (= relative humidity) $p/p_0=\mathrm{RH}$ ($p_0$ is the saturation pressure of water at 24\dC). Here $\Delta l$ is the measured length change of the membrane and $l_0$ the length of the sample at RH=0\%. The behavior observed is typical of adsorption-induced deformations of mesoporous materials and is due to the change of the pressure $P$ of the confined fluid (adsorption stress) in the mesopores \cite{Gor2010}. The shape of the strain at low pressures (before the hysteresis loop related to capillary condensation) may differ and depends on the strength of the solid-fluid interactions \cite{Gor2013}. At higher pressures, when the pores are filled with capillary condensate, the strain changes linearly with $\ln(p/p_0)$. This trend is not specific to pSi and has been observed for all mesoporous materials investigated so far \cite{Amberg1952,  Schappert2014Langmuir, Dolino1996, Gunther2008, Prass2009, Balzer2015, Mogilnikov2002, Dourdain2008, Reichenauer2001, Herman2006, Grosman2015}. In our case, this region corresponds to pressures $p/p_0 \sim 0.83 - 0.93$. The strain for this region is shown in the inset of Fig.~\ref{fig:expt}.

When a pore is filled with fluid, the pressure $P$ that the fluid exerts on a pore wall is given by \cite{Gor2010}:
\begin{equation}
\label{solvation}
P = - \gamma_{sl}/R + P_{L},
\end{equation}
where $\gamma_{sl}$ is the solid-liquid surface energy, $R$ is the pore radius and $P_{L}$ is the Laplace pressure at the liquid meniscus. The latter can be written as (Kelvin-Laplace equation)
\begin{equation}
\label{Laplace}
P_L = \frac{R_g T}{V_l} \ln\left(\frac{p}{p_0}\right)
\end{equation}
where $R_g$ is the gas constant, $T$ is the temperature and $V_l$ is the molar volume of the fluid. Since both the pressure in the pore and the experimentally measured strain of the porous sample as a whole vary linearly with $\ln(p/p_0)$, it is easy to get the pore-load modulus from the following relation $P = M \epsilon_m + C$, where $C$ is a constant, related to the surface energy of the solid-fluid interface. Although the value of $C$ is important for studying the thermodynamics of confined fluids \cite{Gor2014}, it does not affect the present discussion. Fitting of the data shown in the inset of Fig.~\ref{fig:expt} gives $M = 34.5$~GPa.

The key problem is to find the relation between the pore-load modulus $M$ and the elastic constants of the non-porous material: Young's modulus $E$ and Poisson's ratio $\nu$. We assume isotropic elasticity. {Note that by the Young's modulus $E$ we mean the modulus of the pore walls, and not the effective Young's modulus of a pSi sample as a whole ($E_p$), often reported in the pSi literature \cite{Canham2015M}. These two moduli should not be confused, since the latter is a property of a solid-void composite and is a strong function of the porosity \cite{Bellet1996, Magoariec2009}.} 

The deformation of a system with multiple pores can be solved numerically using FEM \cite{Hecht2012}. Since for the considered material the PSD is narrow, the pores are parallel and distributed with a short-range hexagonal order, the pore-load modulus can be approximately calculated from deformation of a two-dimensional hexagonal lattice. The FEM representation of the problem is shown in Fig.~\ref{fig:porous}. The left and bottom surface are constrained in the $x$ and $y$ directions respectively, while the right and top surface move freely. The pressure $P$ is applied in every pore, and the pore load modulus is estimated as the ratio of the applied pressure to the average engineering strain in the vertical direction. Poisson ratio $\nu=0.28$ was used, corresponding to bulk silicon properties \cite{Hopcroft2010}. Systems of two different sizes, 5~$\times$~8 (68 pores) and 10~$\times$~16 (295 pores), were used to check for finite size effects. We found that the results for the larger system differs from the results for the smaller one by 2.5\% at most. To evaluate the role of the lattice geometry we also performed the simulations on a square lattice 14~$\times$~14 (296 pores); the difference between the modulus calculated for square and hexagonal lattices is 4\% at porosities $\phi=35\%$ and 14\% at $\phi=65\%$.

\begin{figure}
\centering
\includegraphics[width=0.95\linewidth]{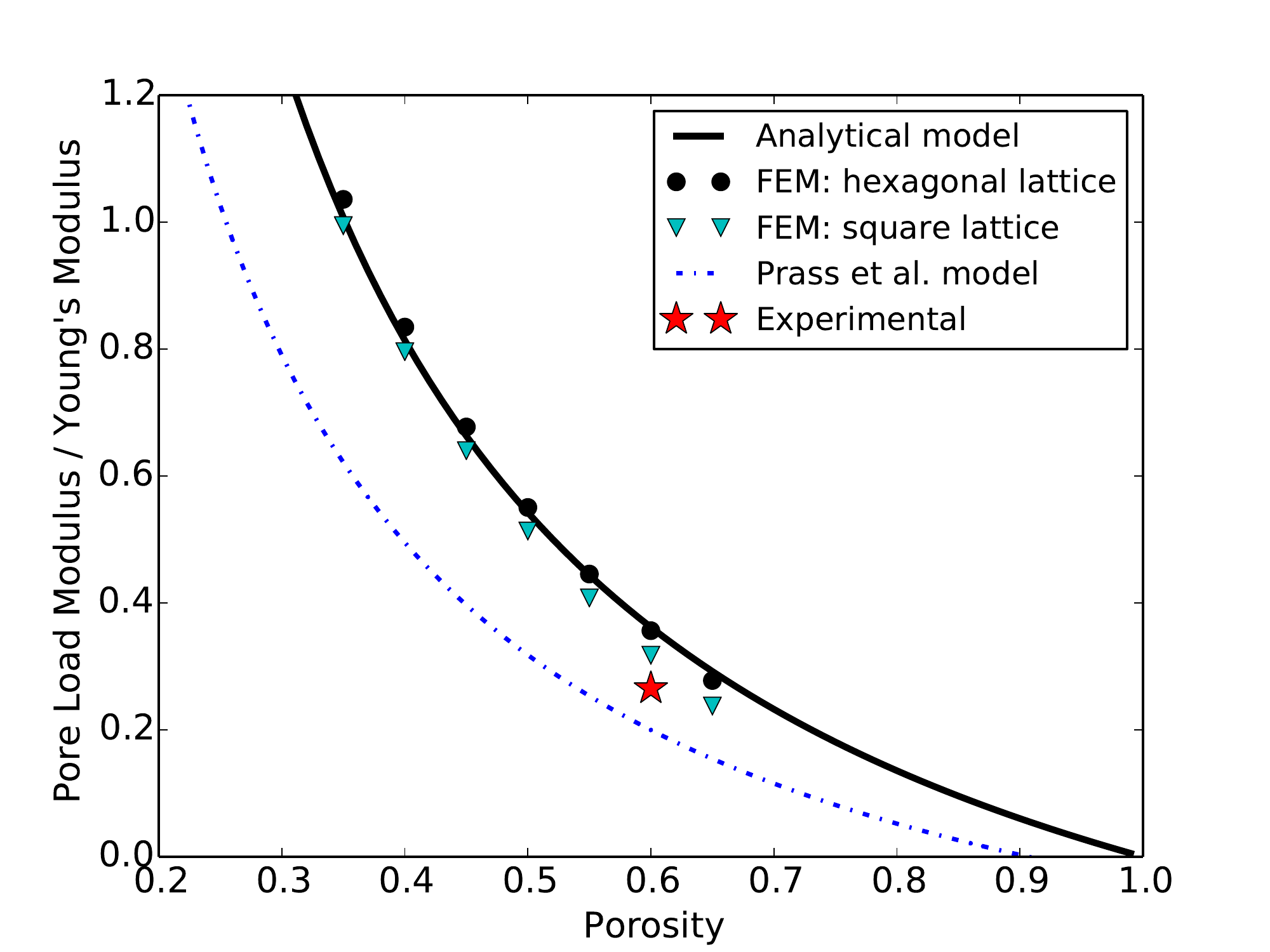}
\caption{Ratio of the pore-load modulus to the Young's modulus of non-porous material as a function of porosity $\phi$ (at $\nu = 0.28$). The results of FEM simulations of deformation of samples with hexagonal and square lattice show that the pore-load modulus weakly depends on the geometrical arrangement of pores, and can be expressed as a function of porosity. The proposed analytical model is in excellent agreement with FEM results. It also shows reasonably good agreement with the pore-load modulus calculated from the experimental data presented in this work. For comparison, the Young's modulus of silicon was taken $\langle$100$\rangle$ $E=130$~GPa \cite{Hopcroft2010}. }
\label{fig:comparison}
\end{figure}

The pore-load modulus calculated from FEM simulations is shown in Fig.~\ref{fig:comparison}. We see that the moduli for different lattice geometries are close, and for a given $E$ and $\nu$ depend mainly on porosity $\phi$ --  the ratio of the volume of the voids to the total volume of the sample. For an analytical estimate of this dependence, we use the following considerations based on classical theory of elasticity. A single cylindrical pore with inner radius $R$ and pressure $P$ induces a purely deviatoric stress in an infinite plate, not leading to any swelling. Swelling is a consequence of the interaction of the pressurized pore with a free surface (where the normal stress must be zero). This is well known for the swelling of a pressurized cylinder with outer radius $R_e$ (Fig.~\ref{fig:porous}) where the engineering strain is calculated from the increase of $R_e$ \cite{Timoshenko1970, Landau1986}:
\begin{equation}
\label{strain-ext}
\epsilon_e = \frac{u(R_e)}{R_e} = \frac{2 \xi (1 - \nu^2)}{E (1 - \xi )} P,
\end{equation}
where $\xi \equiv R^2/R_e^2$. In a large plate containing many pores arranged e.g., on a hexagonal or a square lattice, symmetry implies that the stress must vanish at half the distance between neighboring pores. Hence, the forces compensate at this point making it actually equivalent to a free surface. This suggests that the dilatational strain in the plate with many pores can be approximated by Eq. \ref{strain-ext}, where $R_e$ is taken as half the nearest neighbor distance in the pore lattice. The remaining question is how to relate $\xi$ to $\phi$. One possibility would be to use the lattice geometry to relate nearest neighbor distance to pore fraction, although this neglects the mechanical influence of bits of material outside the symmetry planes. This would give the relation $\xi = (2 \sqrt{3}/\pi) \phi $ for the hexagonal and $\xi = (4/\pi) \phi  $ for the square lattice, respectively. The best fit to the FEM data, however, is obtained with $\xi = \phi$ and, in the absence of a more detailed analytical theory, we use this and write the pore-load modulus approximately as 
\begin{equation}
\label{Me}
M_e = \frac{E}{2(1-\nu^2)} \left(\phi^{-1} - 1\right) .
\end{equation}
Note that Eq.~\ref{Me} gives the the modulus in terms of the basic material properties $E$, $\nu$ and porosity $\phi$.

The results for the pore-load modulus of porous silicon are shown in Fig.~\ref{fig:comparison}. The solid line shows the analytical model Eq. \ref{Me}, which is in excellent agreement with the modulus $M$ derived from our FEM calculations for a porous body. Fig.~\ref{fig:comparison} also shows that our model predicts a higher pore-load modulus than the model by Prass et al. \cite{Prass2009}, especially at porosities below $\sim$~60\%. Such deviation is related to the thin wall approximation used in \cite{Prass2009}, which fails when the porosity decreases.

Comparison of the model to our experimental data (Fig.~\ref{fig:comparison}) shows that if we assume that the Young's modulus of pSi walls equals to the bulk $\langle$100$\rangle$ value, the experimental data is close to the theoretical predictions, yet slightly below them. This deviation can be due to the irregularity of the pSi structure. Both the deviation of pore sizes from the mean size (i.e. PSD) and deviation of the pore arrangement from a perfect lattice lead to presence of thinner pore walls, which reduces the pore-load modulus of the structure. Also, the Young's modulus of the walls can be somewhat lower than the modulus for bulk silicon; applying our model to calculate the Young's modulus from our experimental value of $M$, we get $E = 95$~GPa. Note, however, that here the deviation of Young's modulus of the pore walls from the Young's modulus of bulk silicon is only 27\% and not a factor of five as reported in \cite{Grosman2015}. 

The system of hexagonally-ordered parallel cylindrical pores is not unique to mesoporous silicon: a number of templated mesoporous materials have similar morphology, among them MCM-41 and SBA-15 silica, and silica monoliths with hierarchical pores, synthesized recently \cite{Balzer2015}. The model for the pore-load modulus proposed in this work can be applied to develop quantitative models of adsorption-induced deformation of these materials. It should provide better predictions than the earlier model \cite{Prass2009} at porosities $60\%$ and below, typical for mesoporous silica. 

To summarize, we have presented an experimental study of water adsorption-induced deformations of a monolithic, mesoporous silicon membrane, and extracted the pore-load modulus. We have proposed an analytical model which relates the pore-load modulus to the porosity and to the elastic properties of bulk silicon, and justified it by comparing with FEM simulations. We have found good agreement between our experimental data and the predictions of our model, with the Young's modulus of the pore walls slightly lower than the bulk value for silicon. Our model can be applied to a large class of materials with morphologies similar to mesoporous silicon. Moreover, our findings suggest that liquid condensation experiments allow one to access the elastic constants of a mesoporous medium.

This research was performed while one of the authors (G.G.) held a National Research Council Research Associateship Award at Naval Research Laboratory. G.G. thanks John Michopoulos for insightful discussions of the FEM representation of the problem.  The work of G.G. and N.B. was funded by the Office of Naval Research through the Naval Research Laboratory's basic research program. P.H. acknowledges support by the German research foundation (DFG) within the collaborative research initiative ``Tailor-made Multi-Scale Materials Systems'' (SFB 986, project area B, Hamburg).

\end{document}